 \documentclass[manuscript]{aastex}

\slugcomment{}

\shorttitle{Morphology of Seyfert Galaxies}
\shortauthors{Chen et al.}

\begin{document}

\title{Morphology of Seyfert Galaxies}

\author{Yen-Chen Chen and Chorng-Yuan Hwang}
\affil{Graduate Institute of Astronomy, National Central University, Chung-Li 32054, Taiwan}
\email{m1029006@gm.astro.ncu.edu.tw \& hwangcy@gm.astro.ncu.edu.tw}



\begin{abstract}
We probed the relation between properties of Seyfert nuclei and morphology of their host galaxies. We selected Seyfert galaxies from the Sloan Digital Sky Survey with redshifts less 0.2 identified by the V\'{e}ron Catalog (13th). We used the ``{\it{FracDev}}'' parameter from SDSS galaxy fitting models to represent the bulge fractions of the Seyfert host galaxies. We found that the host galaxies of Seyfert 1 and Seyfert 2 are dominated by large bulge fractions, and Seyfert 2 galaxies are more likely to be located in disk galaxies whereas most of the Seyfert 1 galaxies are located in bulge-dominant galaxies. These results indicate that the types of AGNs are related to their host galaxies and can not be explained by the traditional unification model of Seyfert galaxies. 
\end{abstract}





\section{Introduction}

Active galactic nucleus (AGN) galaxies are galaxies showing strong activity in their nuclei. An AGN is believed to consist of a central engine with an accretion disk and a supermassive black hole embedded in an opaque torus of dust \citep{Rowan77,Antonucci85}. The phenomena of AGN are widely explained by the accretion process of the central supermassive black hole, which would release huge energy \citep{Rees84}. Although AGNs have various types, they are considered to be similar objects and can be explained by the AGN unification model \citep[e.g.,][]{Antonucci93, Urry95}.

One of particular AGN types is called Seyfert galaxies \citep{Seyfert43}. \citet{Seyfert43} found that there were some spiral galaxies with bright central nuclei, and the spectra of the nuclei showed strong emission lines. \citet{Khachikian74} classified Seyfert galaxies into two subclasses according to the line widths of Balmer lines and [OIII] forbidden lines; Seyfert 1 galaxies have broader Balmer lines than the forbidden lines whereas Seyfert 2 galaxies have the same line widths of Balmer lines and the forbidden lines with line width ranging from 500 $\mathrm{km~s^{-1}}$ to $1000~\mathrm{km~s^{-1}}$. \citet{Osterbrock77,Osterbrock81} further divided the Seyfert galaxies into subclasses of Seyfert 1.2, Seyfert 1.5, Seyfert 1.8 and Seyfert 1.9 depending on the appearance of H$\beta$ emission line; Seyfert 1.2 have strong broad H$\beta$ component while Seyfert 1.8 have very weak broad H$\beta$ component in the optical spectra. Seyfert 2 galaxies usually have a high ratio of [OIII]/H$\beta$; the empirical criterion for Seyfert 2 is [OIII]/H$\beta$ $\geq3$  \citep{Shuder81,Veilleux87}. \citet{Baldwin81} introduced a method, called BPT diagram, to divide star-forming galaxies and Seyfert 2 galaxies depending on the ratios of emission lines. Nowadays, the BPT diagrams with different dividing lines derived from theoretical and empirical methods are widely used in diagnosing Seyfert 2 galaxies \citep[e.g.,][]{Kauffmann03,Kewley06,Schawinski07}.

According to unification model \citep{Antonucci93, Urry95}, different types of AGNs are caused by different viewing angles. For example, Seyfert 1 galaxies are viewed face-on relative to the accretion disk and torus whereas Seyfert 2 galaxies are viewed edge-on \citep{Antonucci93}. \citet{Antonucci93} found that NGC 1068 showed broad emission lines in polarized spectroscopic observations while it was considered as a Seyfert 2 galaxies in traditional optical spectral observations. This result strongly supports the unification model of Seyfert 1 and Seyfert 2 galaxies.

If the Seyfert galaxies are merely due to different viewing angles relative to torus, different types of Seyferts should be independent of their host galaxies. However, \citet{Xanthopoulos96} found that Seyfert galaxies tend to be in the S0 and Sa galaxies with 27 Seyferts selected from the V\'{e}ron catalogue (1985). \citet{Maiolino97} found that Seyfert 2 and Seyfert 1 galaxies have similar CO distributions but the host galaxies of Seyfert 2 seem to have more asymmetric morphology than those of Seyfert 1. \citet{Malkan98} found that he median in morphological classes for Seyfert 1 galaxies is Sa and that for Seyfert 2 galaxies is Sb from the WPFC2 images of 256 nearest active galaxies; the subsample of these Seyfert galaxies selected from 12~$\mu$m emission also show a similar trend \citep{Malkan98}. \citet{Hunt99} found that the median type of Seyfert 1 galaxies is Sa while that of Seyfert 2 galaxies is Sab galaxies using 891 galaxies from a 12~$\mu$m galaxy sample. \citet{Koulouridis06} found that there are significantly higher fraction of Seyfert 2 galaxies with a neighbor within 75 $h^{-1}$ kpc than Seyfert 1 galaxies; the authors also found that Seyfert 1 galaxies prefer to be in more dense galaxy regions than Seyfert 2 galaxies do in large scale environments. The above results seem to conflict with the traditional unification model, suggesting that the orientation in unification model is not the only reason for different Seyfert-type galaxies.


In order to understand whether the different types of Seyfert galaxies are related to their host galaxies, we studied the host galaxy morphology distributions of a sample of selected Seyfert galaxies. This paper is organized as follows. In Section 2, we describe our sample selection. In Section 3, we present the distributions of host galaxy morphology for our Seyfert samples. We discuss and summarize our results in Section 4. In this paper, we used $H_{0}$=70~km~s$^{-1}$~Mpc$^{-1}$, $\Omega_{m}=0.3$, $\Lambda_{0}=0.7$, $q_{0}=-0.55$, $k=0.00$.

\section{Sample Selection}


We used two different methods to select our Seyfert samples from the Sloan Digital Sky Survey (SDSS) to avoid possible selection biases. For the first method of selection, we chose our Seyfert galaxies identified by the V\'{e}ron Catalog (13th) in the SDSS data. We selected our Seyferts from the Table\_AGN of the V\'{e}ron 13th Catalog \citep{Veron10}. The table comprised Seyfert 1s, Seyfert 2s, and LINER with $\mathrm{M}_B$ fainter than -22.25. The Seyfert galaxies in this catalogue were divided into six subclasses \citep{Osterbrock77,Osterbrock81} : Seyfert 1.0, Seyfert 1.2, Seyfert 1.5, Seyfert 1.8, Seyfert 1.9 and Seyfert 2 depending on the appearance of their Balmer lines. Seyfert 1 have broad Balmer and other emission lines while Seyfert 2 have the narrow Blamer and forbidden lines. We only selected Seyfert 1.0 galaxies as our Seyfert 1 samples and Seyfert 2.0 as our Seyfert 2 samples to avoid confusion. We also constrained the redshift range of the Seyfert galaxies from 0 to 0.2 to include some important emission lines within the spectral range. There are 5009 Seyfert 1 galaxies and 4204 Seyfert 2 galaxies after these selections. We extracted the host galaxies of the Seyferts from the Sloan Digital Sky Survey (SDSS) and used the observation data from SDSS Data Release 10 \citep{Ahn14}. We selected SDSS sources that were classified as ``GALAXY''  in photometry within a 3'' radius of our Seyfert samples. We only considered the sources that contain both photometric and spectroscopic information. Finally, we have 4078 Seyfert 1 galaxies and 3422 Seyfert 2 galaxies.




\section{Host Galaxy Morphology}

\begin{figure}[htbp]
\begin{center}
\includegraphics[width=0.9\textwidth]{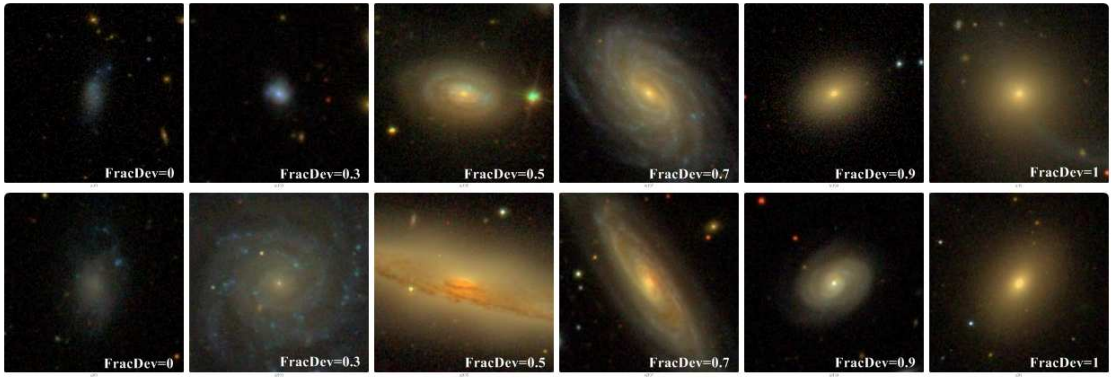}
\caption{Example images of Seyfert galaxies as a function of {\it{FracDev}}. Top row: Seyfert 1 samples. Bottom row: Seyfert 2 samples.}
\label{s1s2ima}
\end{center}
\end{figure}

To give a quantity assessment of galaxy morphology, we used the SDSS parameter, {\it{FracDev}}, to represent the bulge contributions of our Seyfert galaxies.
\begin{displaymath}
F_{\mathrm{composite}}=FracDev~F_{\mathrm{dev}}-(1-FracDev)~F_{\mathrm{exp}}
\end{displaymath}
where $F_{\mathrm{composite}}$ is the composite model of galaxy flux, $F_{\mathrm{dev}}$ is the best fitting of the de Vaucouleurs flux of the galaxy, and $F_{\mathrm{exp}}$ is the best fitting of the exponential disk flux of the galaxy. The {\it{FracDev}} parameter is a coefficient of the de Vaucouleurs bulge term in the SDSS galaxy fitting and has a range from 0 to 1 \citep{Bernardi06}. If the value of {\it{FracDev}} is close to 1, the galaxy is bulge-dominant \citep{Kuehn05, Vincent05}. From the {\it{FracDev}} values, we can determine the bulge contributions in the host galaxies. There is an example of color images of our Seyfert galaxies as a function of {\it{FracDev}} in Fig.~\ref{s1s2ima}. We used {\it{FracDev\_{r}}} expressing the coefficient fitting by {\it{r}}-band data for better quality. All {\it{FracDev}} in this paper present the coefficient fitting by {\it{r}}-band data.

\begin{figure}
\begin{center}
\includegraphics[width=\columnwidth]{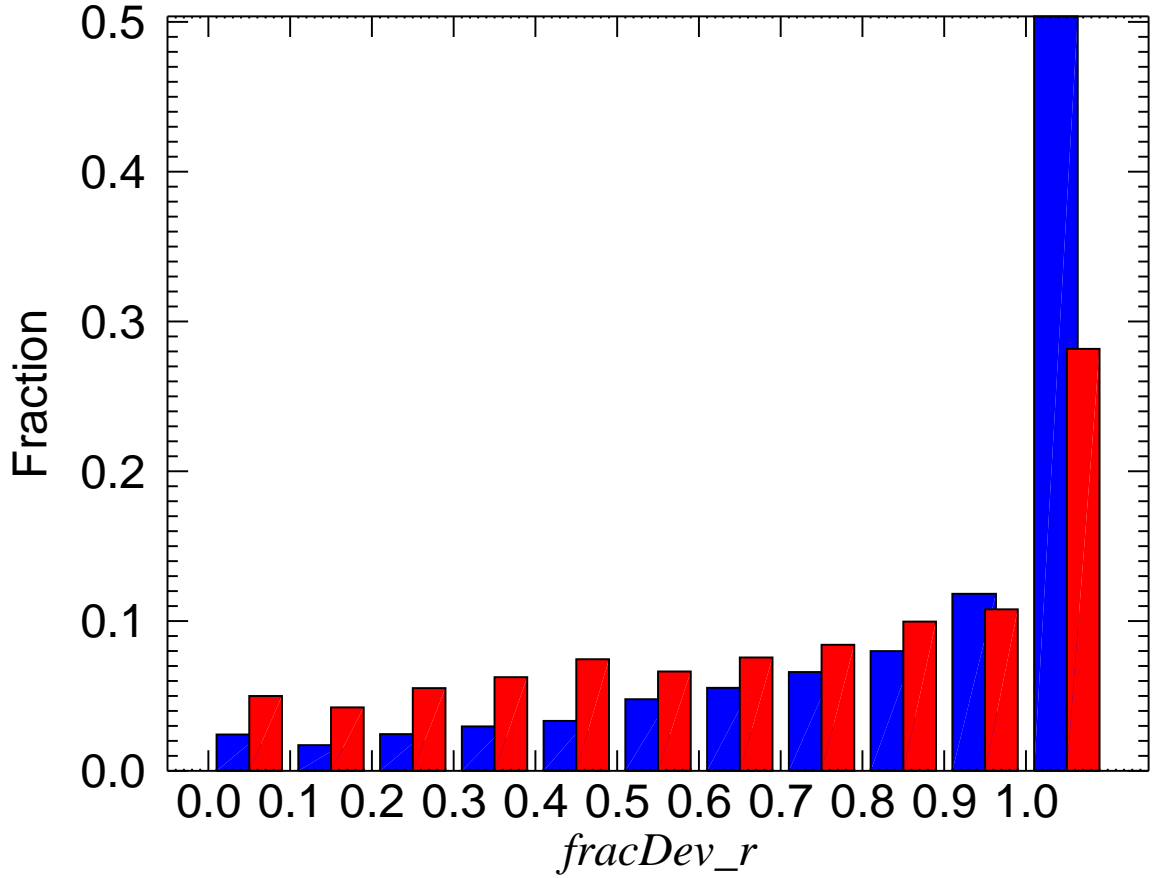}
\caption{Host galaxy morphology distributions of Seyfert galaxies identified by the V\'{e}ron Catalog in the SDSS data. Blue color represents the Seyfert 1 galaxies and red color represents the Seyfert 2 galaxies.}
\label{s1s2fd}
\end{center}
\end{figure}

Fig.~\ref{s1s2fd} shows the host galaxy morphology distributions of our Seyfert galaxies. We found that Seyfert 1 and Seyfert 2 galaxies show different distributions. We did a Kolmogorov-Smirnov (K-S) test for the {\it{FracDev}} distributions of Seyfert 1 and Seyfert 2 galaxies. The probability that the {\it{FracDev}} distributions of Seyfert 1 and Seyfert 2 galaxies are drawn from the same population is $P = 0.00$ with K-S statistic (D) $=0.26$. In other words, the host galaxies of these two Seyferts have different morphology distributions and the different distributions in host morphology between the Seyfert 1 and Seyfert 2 galaxies are not biased by the different selection effect. We also compared the fractions of Seyfert galaxies for both {\it{FracDev}}~=~1 and {\it{FracDev}}~$<$~1. The fraction of Seyfert 1 with {\it{FracDev}}~=~1 is 50\% but for Seyfert 2 galaxies the fraction is 28\%. This indicates that the Seyfert 2 galaxies are more likely to locate in disk galaxies whereas most of the Seyfert 1 galaxies are located in bulge-dominant galaxies. Besides, the result shows that Seyfert 1 and Seyfert 2 galaxies are dominated by galaxies with {\it{FracDev}}~=~1. In other words, there are many Seyfert galaxies located in elliptical type galaxies, in contradiction to early studies \citep{Adams77, Heckman78}.

\begin{figure}
\begin{center}
\includegraphics[width=\columnwidth]{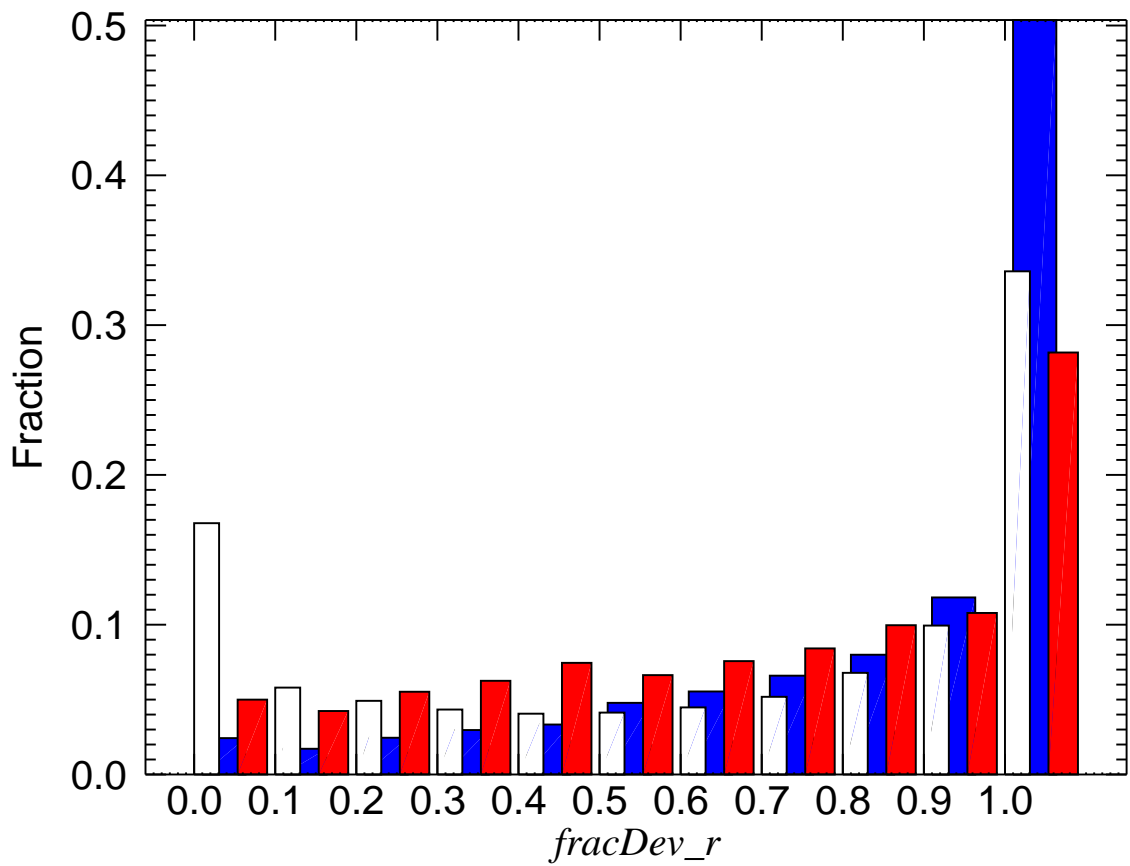}
\caption{Host galaxy morphology distributions of our Seyfert galaxies and all galaxies in SDSS with redshifts less than 0.2 and $M_{\mathrm{r}} <$ -19. Blue bars represent the Seyfert 1 galaxies, red bars the Seyfert 2 galaxies, and white bars all the SDSS galaxies with z $<$ 0.2 and $M_{\mathrm{r}} <$ -19.}
\label{s1s2ng}
\end{center}
\end{figure}

We compared the morphology distributions of Seyfert galaxies with all galaxies that have redshifts less than 0.2 and $M_{\mathrm{r}} <$ -19 in the SDSS database. Fig.~\ref{s1s2ng} demonstrates that the morphology distribution of the total galaxies is different from those of the Seyfert galaxies. These results show that Seyfert galaxies are more bulge dominant than all galaxies.

\begin{figure}
\begin{center}
\includegraphics[width=\columnwidth]{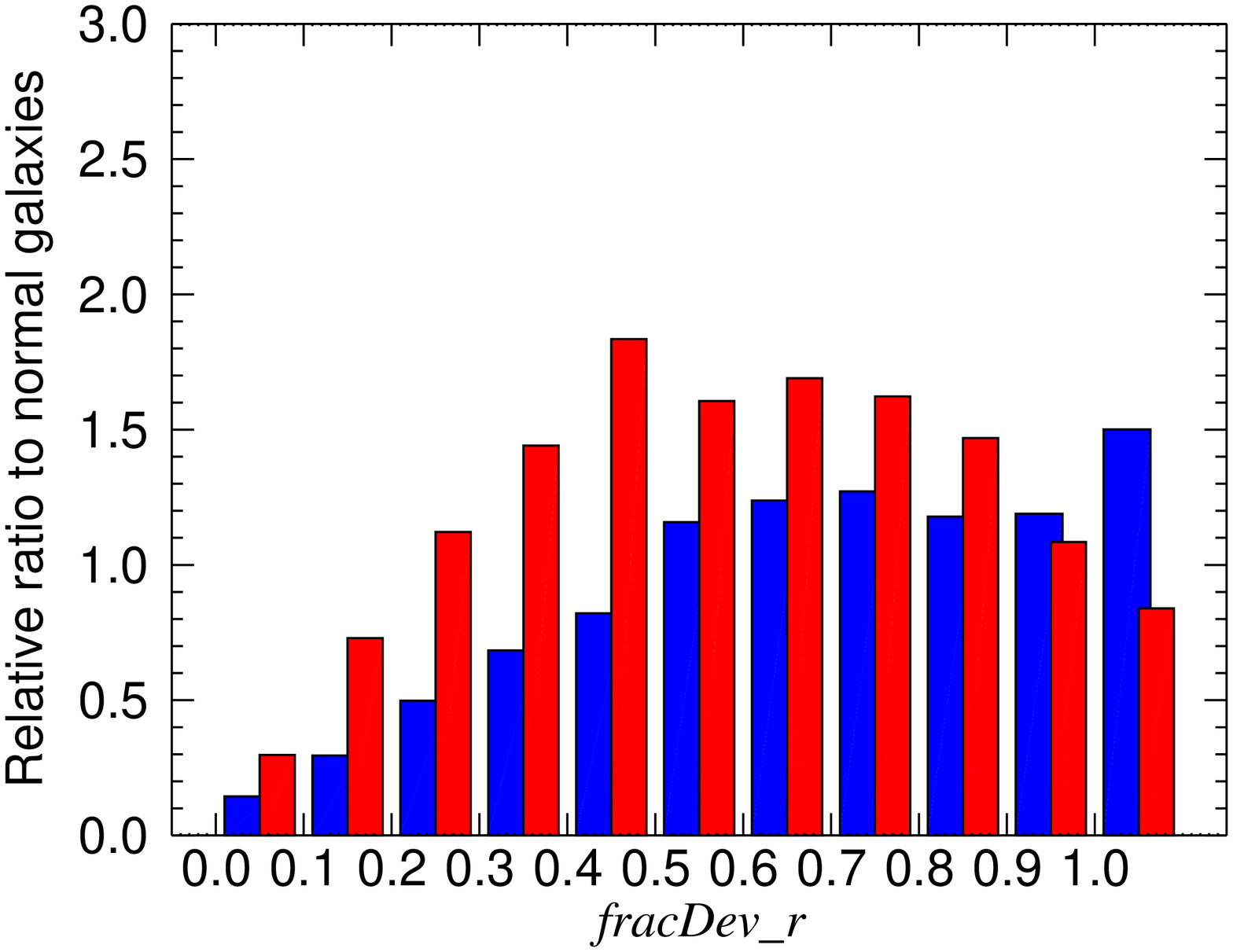}
\caption{Ratios of Seyfert galaxies to all SDSS galaxies with z $<$ 0.2 and $M_{\mathrm{r}} <$ -19. The symbols are the same as in Fig.~\ref{s1s2fd}.}
\label{s1s2dng}
\end{center}
\end{figure}

Furthermore, we plotted the relative ratios of the Seyfert galaxies to the total galaxies with different {\it{FracDev}} values in Fig.~\ref{s1s2dng}. Fig.~\ref{s1s2dng} suggests that the Seyfert 2 galaxies are more likely to emerge in {\it{FracDev}} $\approx$ 0.4 whereas the Seyfert 1 galaxies are more likely to emerge in {\it{FracDev}}~=~1. In other words, the host galaxies of the Seyfert 1 are more likely to be early type galaxies and those of the Seyfert 2 are more likely to be late type galaxies.

\begin{figure}
\begin{center}
\includegraphics[width=\columnwidth]{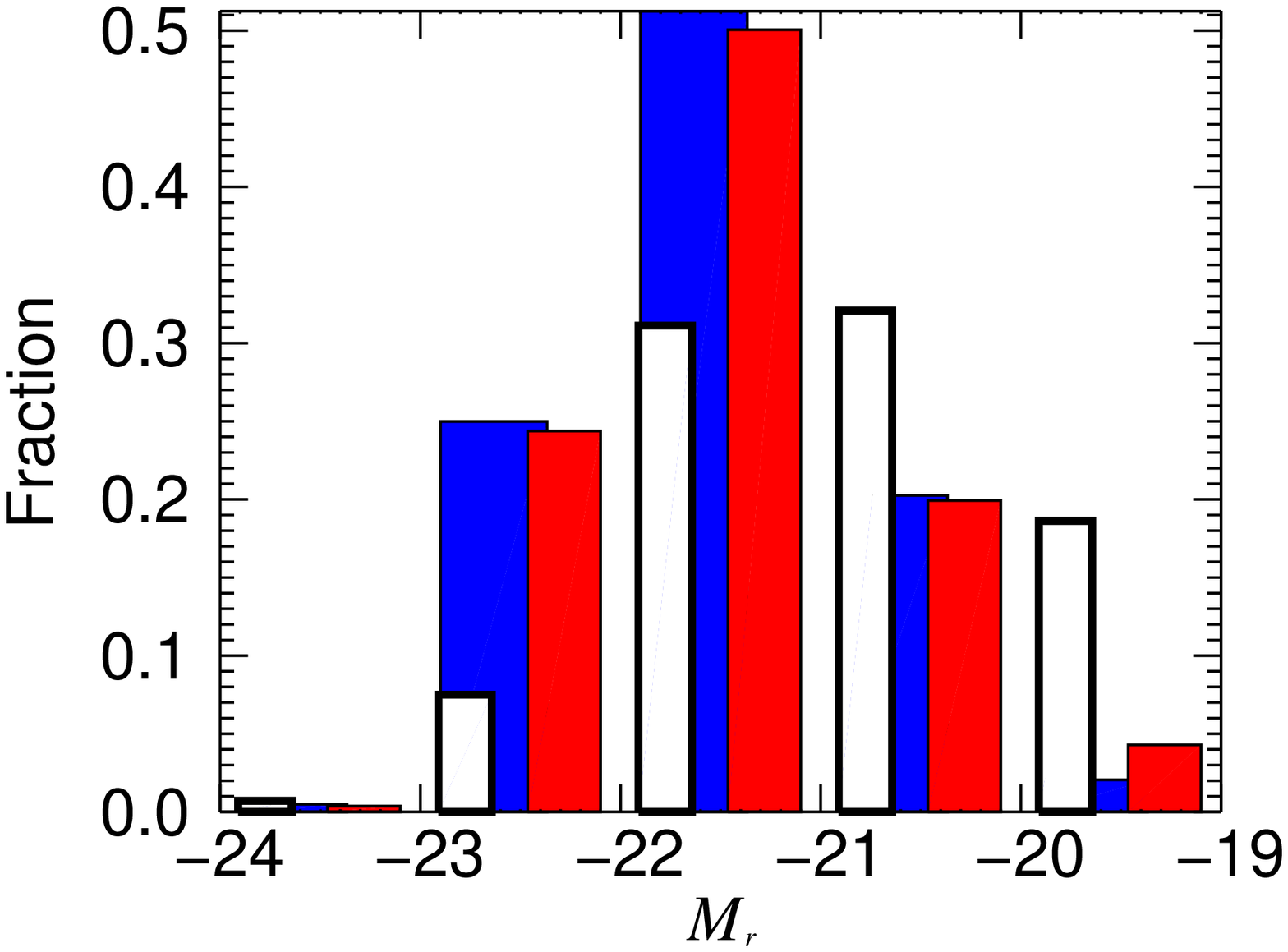}
\caption{Distributions of {\it{r}}-band absolute magnitudes for Seyfert 1, Seyfert 2, and total galaxies. Blue bars represent the Seyfert 1 galaxies, red bars the Seyfert 2 galaxies, and white bars all SDSS galaxies with z $<$ 0.2}
\label{s1s2nMdis}
\end{center}
\end{figure}

\begin{figure}
\begin{center}
\includegraphics[width=\columnwidth]{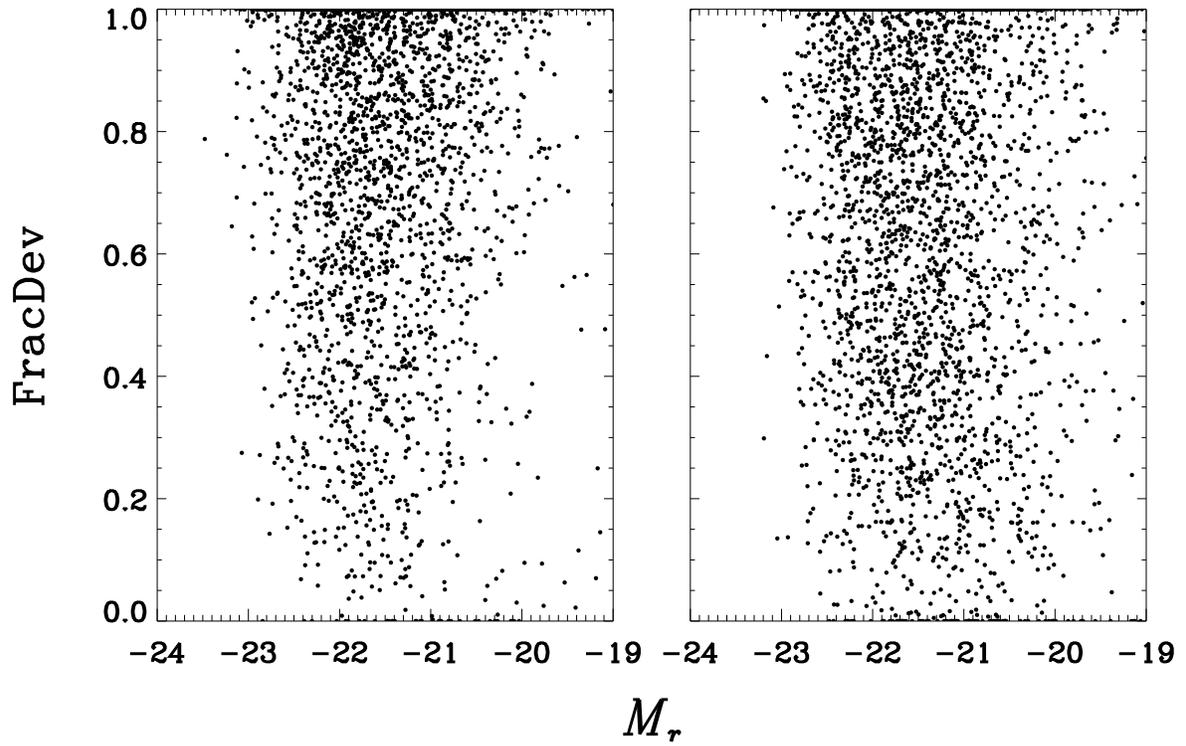}
\caption{Distributions of {\it{r}}-band absolute magnitude versus {\it{FracDev}} for Seyfert 1 and Seyfert 2 galaxies. Left: Seyfert 1 galaxies. Right: Seyfert 2 galaxies. The bin size is one magnitude.}
\label{s1s2MrFra}
\end{center}
\end{figure}

Fig.~\ref{s1s2nMdis} shows the distributions of absolute magnitudes for Seyfert 1, Seyfert 2 galaxies, and total galaxies. The Seyfert 1 and the Seyfert 2 galaxies have similar luminosity distributions, which peak at around absolute magnitudes -22 to -21; on the other hand, the total galaxies have a broader distribution extending to lower luminosities. Fig.~\ref{s1s2MrFra} shows the distributions of $M_{\mathrm{r}}$ and {\it{FracDev}} for our Seyfert galaxies. The {\it{FracDev}} are widely distributed over different $M_{\mathrm{r}}$. In other words, the different morphologies between the Seyfert 1 and the Seyfert 2 galaxies are not related to the absolute magnitudes of their host galaxies.


\begin{figure}
\begin{center}
\includegraphics[width=\columnwidth]{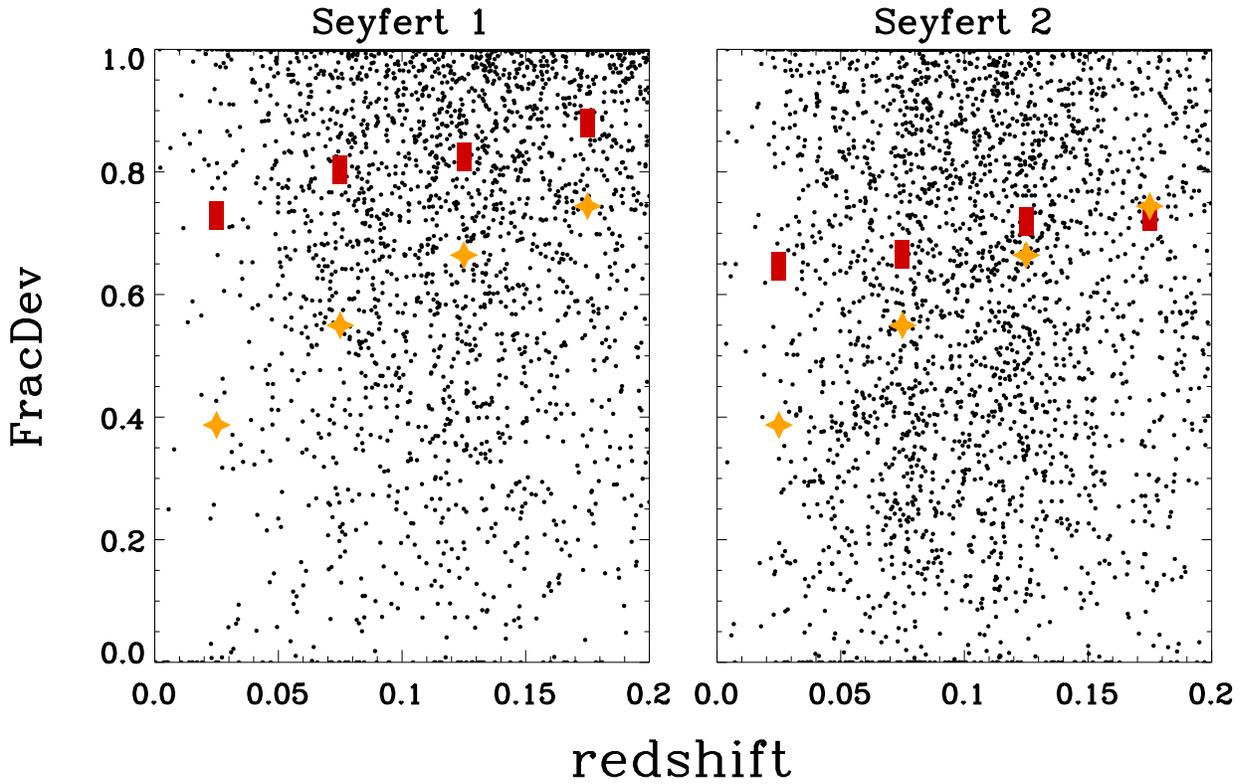}
\caption{Distributions of the {\it{FracDev}} and redshift for Seyfert 1 and Seyfert 2 galaxies. Left: Seyfert 1 galaxies. Right: Seyfert 2 galaxies. Red squares represent the average {\it{FracDev}} in different redshift ranges for the Seyfert galaxies. The  orange stars represent the average {\it{FracDev}} in different redshift ranges for total galaxies, whose {\it{FracDev}}-redshift distribution is too dense to show in the diagram because of the huge number of normal galaxies.}
\label{s1s2nzF}
\end{center}
\end{figure}

We also investigated the effect of redshift on the {\it{FracDev}} for our Seyfert galaxies and total galaxies. Fig.~\ref{s1s2nzF} shows the distributions of redshifts and {\it{FracDev}}; the average {\it{FracDev}} values increase with redshifts for both the Seyfert galaxies and the total galaxies. We noted that at the same redshifts, the average {\it{FracDev}} values for the Seyfert 1 are always larger than the Seyfert 2 galaxies; however, at low redshifts, the Seyfert 2 galaxies do have more sources with small {\it{FracDev}} values. The average {\it{FracDev}} values of the total galaxies are much smaller than those of the Seyferts at low redshifts but approaching to the values of the Seyfert 2 at $z \approx 0.2$. These results suggest that the host galaxies of the Seyfert 1 are more bulge-dominated than those of the Seyfert 2.

\section{Discussion and Summary}

\begin{figure}
\begin{center}
\includegraphics[width=\columnwidth]{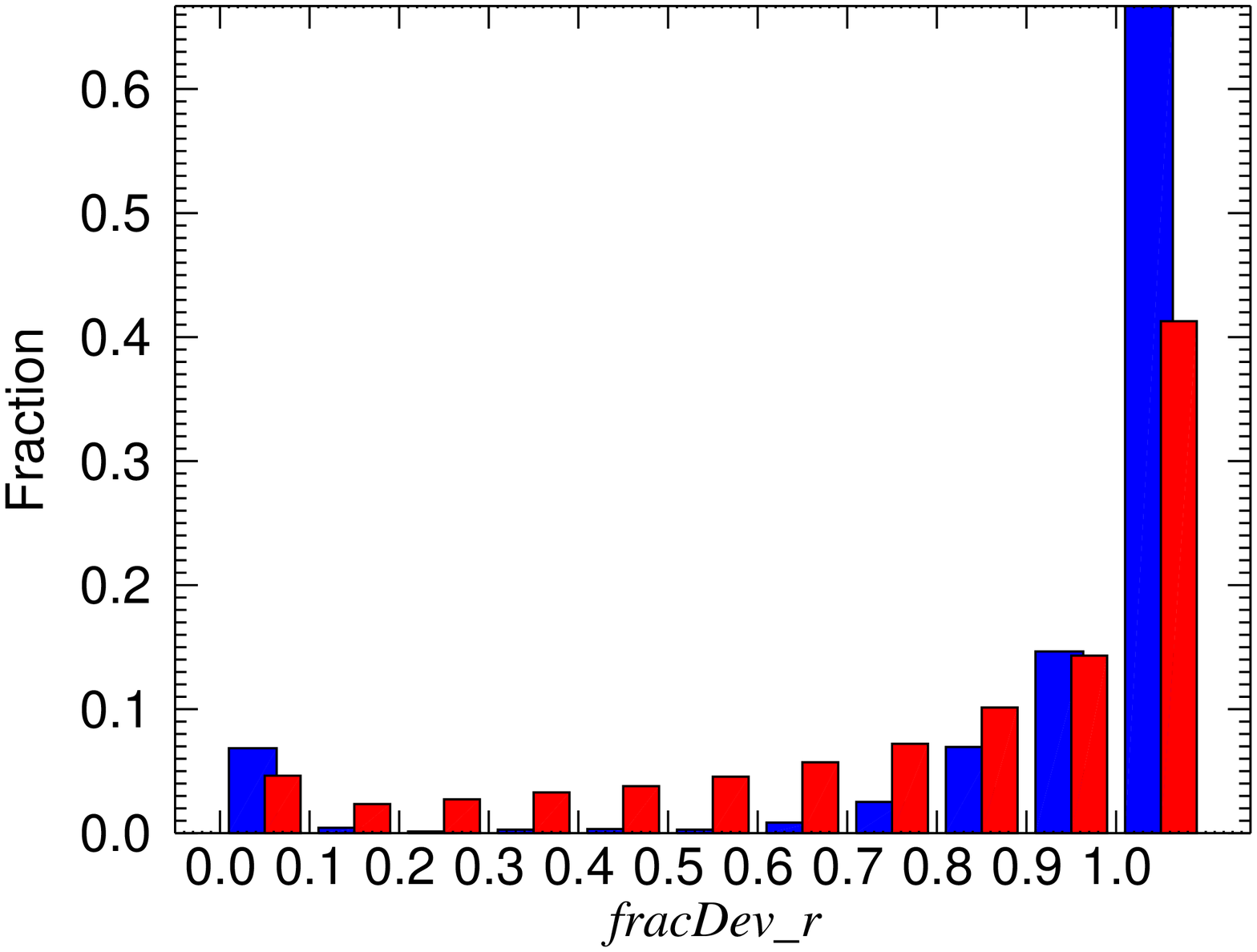}
\caption{Host galaxy morphology distributions of the new selected Seyfert galaxies. Blue color represents the Seyfert 1 galaxies and the red color represents the sample A Seyfert 2 galaxies selected with [OIII]/H$\beta > 3.$}
\label{s1s2cm}
\end{center}
\end{figure}

\begin{figure}
\begin{center}
\includegraphics[width=\columnwidth]{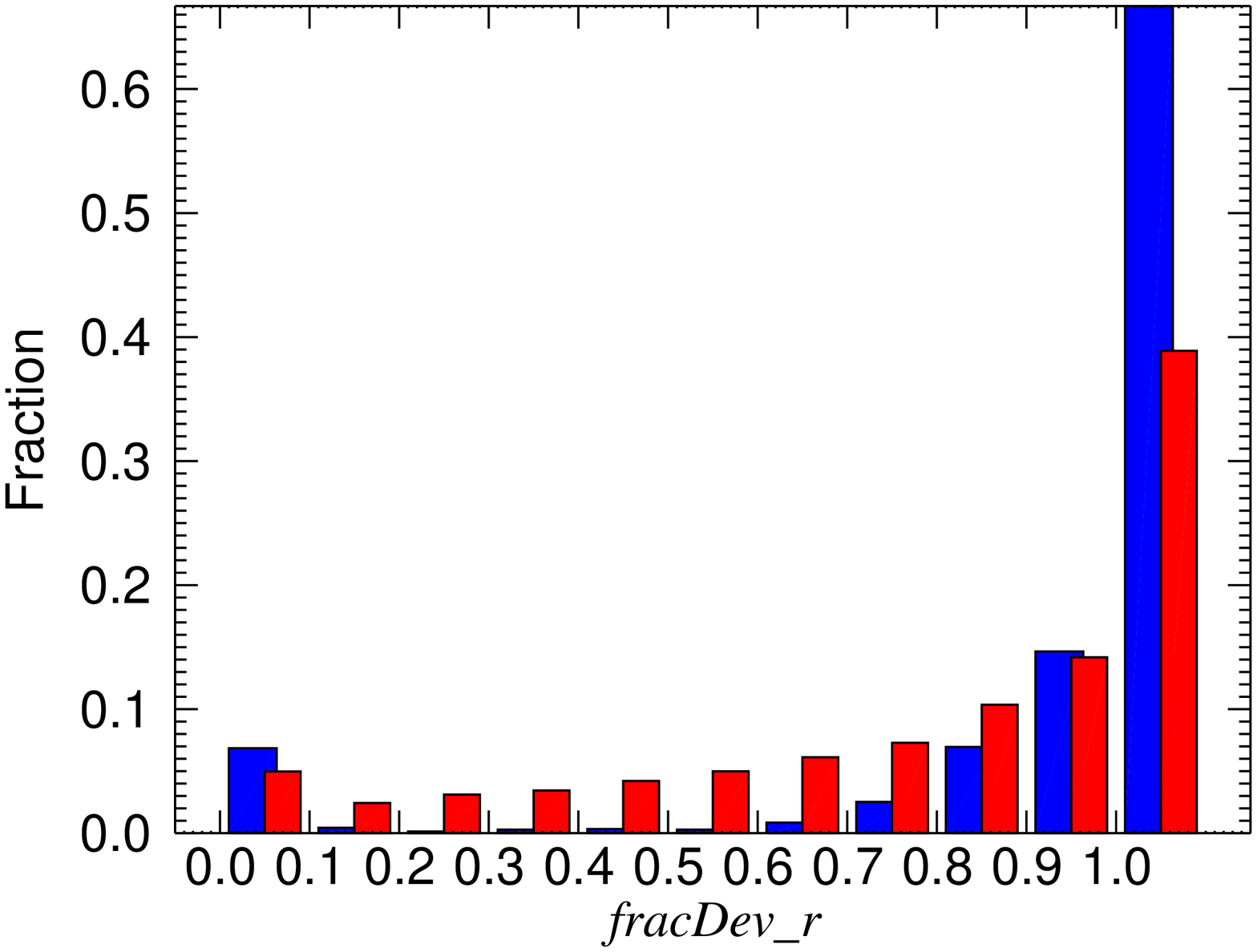}
\caption{Host galaxy morphology distributions of the new selected Seyfert galaxies. Blue color represents the Seyfert 1 galaxies and the red color represents the sample B Seyfert 2 galaxies selected with  the criteria of \citet{Kewley06}.}
\label{s1s2cmkew}
\end{center}
\end{figure}

\subsection{Selection effect}
In order to test whether our results are affected by the selection biases of the V\'{e}ron catalog, we used a different selection method to select a different Seyfert sample for comparison. We selected a new Seyfert samples independently from the SDSS DR10 using typical criteria for Seyfert galaxies without invoking the identification of the V\'{e}ron catalog. We selected the new Seyfert 1 galaxies with the full width at half maximum (FWHM) of the Balmer line $> 1000~\mathrm{km~s^{-1}}$ and H$\beta$/[OIII] $>5$ \citep{Winkler92}. We selected two new samples of  Seyfert 2 with different criteria of line ratios. Sample A of the new Seyfert 2 is selected with [OIII]/H$\beta > 3$ \citep{Shuder81} with the FWHM of the Balmer lines $< 1000~\mathrm{km~s^{-1}} $; sample B of the new Seyfert 2 is selected by the criteria of \citet{Kewley06} with the FWHM of the Balmer lines $< 1000~\mathrm{km~s^{-1}} $. The new Seyfert samples have redshifts between 0 and 0.2, which is the same as the redshift range of our original samples.  We finally have 2102 Seyfert 1 galaxies, 36134 sample A Seyfert 2 galaxies, and 35954 sample B Seyfert 2 galaxies from the new criteria.

Fig.~\ref{s1s2cm} shows the {\it{FracDev}} distributions of the new Seyfert 1 and the sample A Seyfert 2 galaxies. A K-S test shows that the two distributions are drawn from the same population has K-S statistic (D) $=0.37$ and $P = 0.00$.  Fig.~\ref{s1s2cmkew} present the {\it{FracDev}} distributions of the new Seyfert 1 and the sample B Seyfert 2 galaxies, which have a K-S test result with K-S statistic (D) $=0.35$ and $P=0.00$. In other word, we still found that the {\it{FracDev}} distributions of the Seyfert 1 and the Seyfert  2 are completely different even these sources are selected uniformly over SDSS without invoking any V\'{e}ron information. These suggest that our results are very robust and are not affected by the biased of the V\'{e}ron catalogs, which compiled the sources from literature.

\subsection{AGN contributions}

\begin{figure}
\begin{center}
\includegraphics[width=\columnwidth]{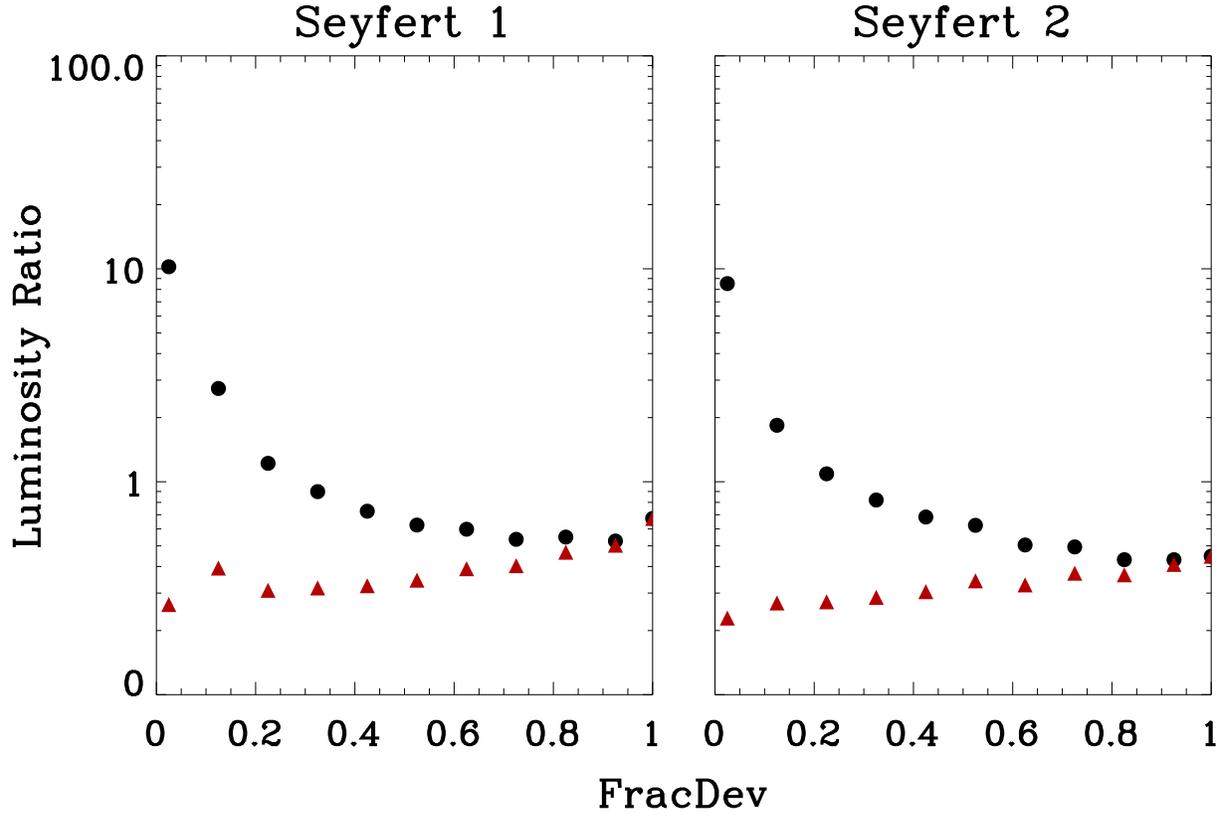}
\caption{Ratios of AGN contributions to the luminosities of their host galaxies and bulges. Left: Seyfert 1 galaxies. Right: Seyfert 2 galaxies. Black dots represent the averaged ratios of AGN luminosities to their bulge luminosities within 0.1 intervals of the {\it{FracDev}} values. Red triangles represent the averaged ratios of AGN luminosities to their host galaxy luminosities within 0.1 intervals of the {\it{FracDev}} values .}
\label{s1s2con}
\end{center}
\end{figure}

Fig.~\ref{s1s2fd} shows that Seyfert 1 and Seyfert 2  galaxies have completely different morphology distributions. We note that Seyfert 1 and Seyfert 2 galaxies have different AGN strengths, which might cause some biases about the morphological fitting. To quantify the effects of the AGN strengths on the morphological fitting, we considered the contributions of the AGNs using the SDSS spectra. Because the SDSS spetra aperture is 3'', the flux inside the aperture should be dominant by AGNs. We could use the fluxes of the spectra to roughly estimate the contributions of AGNs. SpectroFlux\_{\it{r}} is a parameter from the SDSS database and is value of spectrum flux in {\it{r}}-band filter.  We used the parameter to be an indicator of the AGNs contributions.  Fig.~\ref{s1s2con} are the result of average AGNs contributions to their host galaxies in different {\it{FracDev}} ranges. We found that both Seyfert 1 and Seyfert 2 galaxies have similar trends in the ratios of AGNs contributions to their bulges and to their host galaxies. The average ratios of the AGNs to the bulges declined with {\it{FracDev}} and the average ratios of the AGNs to the host galaxies increase with {\it{FracDev}}. If the {\it{FracDev}} fitting was significantly affected by AGNs and different Seyferts have different effects, we should find different trends in the ratios for Seyfert 1 and Seyfert 2 galaxies. Our results of AGNs contributions to the host galaxies suggest that AGNs contributions are unrelated to the results of the host galaxy morphology distributions.

\subsection{Host Galaxy Morphology}

We examined the relation between the {\it{FracDev}} and de Vaucouleurs T value for a subset of our Seyfert galaxies. We constrained the redshifts of the sources to be less than 0.05 and searched for the de Vaucouleurs value of their host galaxies from NASA Extragalactic Database (NED). There are 81 Seyfert 1 galaxies and 127 Seyfert 2 galaxies with available de Vaucouleurs T values. We found that there is a good correlation between the T value and the {\it{FracDev}}, which is consistent with \citet{Oh13}, who found a very good correlation between the {\it{FracDev}} and the Hubble types for 7429 nearby normal galaxies. \citet{Oh13} also showed that the average {\it{FracDev}} is $\approx$ 1 for elliptical and S0 galaxies. These results suggest that the relation between the {\it{FracDev}} and the de Vaucouleurs T values for our Seyfert samples are similar to those of normal galaxies; this similarity shows that our {\it{FracDev}} values are not significantly affected by the central AGNs and the {\it{FracDev}} can be used as an indicator of the host  galaxy morphology.

The difference in the morphology distributions of the Seyfert 1 and the Seyfert 2 galaxies might be biased by the magnitudes of our Seyfert samples. Fig.~\ref{s1s2nMdis} shows that the Seyfert 1 and Seyfert 2 galaxies have similar luminosity distributions toward bright end. Bright galaxies usually have larger fractions of light in their bulges whereas faint galaxies have most of their light in disks \citep[e.g.,][]{Tasca11}. However, Fig.~\ref{s1s2dng} shows that the Seyfert 2 galaxies have relative high ratios at medium {\it{FracDev}} values than the Seyfert 1 galaxies. This indicates that although the galaxy luminosities might affect the morphology distributions of the Seyfert galaxies, it can not be the main cause for the different distributions of the Seyfert 1 and the Seyfert 2. Besides, we noticed that the Seyfert 1 and Seyfert 2 galaxies have similar distributions of r-band absolute magnitude indicating that the Seyfert 1 and Seyfert 2 galaxies have similar stellar mass distributions in their host galaxies. This result suggests that the stellar mass in the different types of Seyfert galaxies is independent of the unification model.

To examine the influence of galaxy luminosity on the morphology, we chose sub-samples from our Seyfert 1 and Seyfert 2 with -22 $< M_{\mathrm{r}} <$ -21 and compared the {\it{FracDev}} distributions for the sub-sample Seyfert 1 and Seyfert 2. The results of the Kolmogorov-Smirnov test on {\it{FracDev}} distributions for the sub-sample Seyfert 1 and Seyfert 2 have probability that the two distributions are drawn from same population $P = 0.00$ with K-S statistic (D)$=0.26$, indicating that the Seyfert 1 and the Seyfert 2 still have different {\it{FracDev}} distributions. These results suggest that the galaxy luminosity does not cause the morphology difference between Seyfert 1 and Seyfert 2.

Traditional unification model \citep{Antonucci93} suggests that the different types of AGNs are due to different observing angles relative to the torus. The unification model suggests that Seyfert 1 and Seyfert 2 galaxies are intrinsically similar objects with different viewing angles. Seyfert 1 galaxies are viewed from face-on to the accretion disk without obscuring, whereas Seyfert 2 galaxies are viewed from edge-on and obscured by the torus. However, our results (Fig.~\ref{s1s2fd}) indicate that the morphology distributions of the host galaxies of Seyfert 1 and Seyfert 2 are different. This suggests that the differences between the Seyfert 1 and the Seyfert 2 are not only caused by the viewing angles but also might be related to their host galaxy morphologies.

In the early study of the host galaxies of AGNs, people found few Seyfert AGNs residing in elliptical galaxies \citep{Adams77, Moles95}. For example, \citet{Moles95} considered AGNs that were classified as Seyfert or LINER in V\'{e}ron catalogue (1991) with morphological information in the Third Reference Catalogue of Bright Galaxies (RC3) and found that only 5\% of their samples were elliptical galaxies. However, some other studies found AGNs residing in early-type galaxies; \citet{Xanthopoulos96} found that the Hubble types of Seyfert galaxies tend to be S0 and Sa; \citet{Ho97} used spectroscopic to study the emission line in the central region of 486 nearby galaxies and found that spectroscopic AGN are dominant in early-type galaxies;  \citet{Schade00} found that there are more than half of their X-ray selected AGNs residing in E/S0 type galaxies. These results suggested that the host galaxy  morphologies of AGNs might prefer to be bulge-dominant galaxies.

The morphology difference of the host galaxies between Seyfert 1 and Seyfert 2 was less clear. \citet{Sorrentino06} found that among 624 Seyfert 1 galaxies, 76\% of the host galaxies of Seyfert 1 are early types and 9.7\% are late types; among 925 Seyfert 2 galaxies, the ratio of late-type galaxies for Seyfert 2 galaxies is 26.9\% and that of early-type galaxies is 56.8\%. \citet{Slavcheva11} found that their Seyfert 1 galaxies prefer to locate in early type galaxies than Seyfert 2 galaxies do. On the other hand, Seyfert 2 galaxies were more frequently found in late type galaxies \citep{Koulouridis06}. Our Seyfert samples are dominated by {\it{FracDev}}=1 for both Seyfert 1 and Seyfert 2. This indicates that most of the Seyfert galaxies are bulge-dominant.  We checked the host galaxy morphology of our Seyferts by using the results of the {\it{Galaxy Zoo Project}} \citep{Lintott08}. There are 2889 Seyfert 1 galaxies and 2826 Seyfert 2 galaxies with available information from the Galaxy Zoo. We used the voted data from the Galaxy Zoo to check the morphology of the galaxies that have {\it{FracDev}}~=~1 \citep{Lintott11}. We assumed that a galaxy with a voted-elliptical fraction $>$ 50\% has the morphology of an elliptical galaxy. We found that  66\% of the Seyfert 1 galaxies with {\it{FracDev}}~=~1 show elliptical-like morphology, and 68\% of the Seyfert 2 with {\it{FracDev}}~=~1 show elliptical-like morphology. Besides, the ratio of elliptical-like galaxies for all Seyfert 1 with different {\it{FracDev}} values is 30\% and that for Seyfert 2 is 19\% from the voting rates of the Galaxy Zoo. Our results are similar to previous studies of \citet{Sorrentino06} and \citet{Slavcheva11}. Furthermore, we assumed that a spiral-like host for a galaxies with voted-spiral fraction $>50\%$ and {\it FracDev} $<1$. We have 20\% spiral host in the Seyfert 1 galaxies and 30\% in the Sefyert 2 galaxies. \citet{Villarroel17} shows that the spiral host of Type-1 AGN is 20\% and Type-2 is 44\%. Our results of spiral host galaxies of Seyferts show that the Seyfert 2 galaxies have more spiral host than Seyfert 1 galaxies and agree with that of \citet{Villarroel17}.

We also investigated whether the morphology distributions of the Seyfert galaxies are affected by redshifts. We selected sub-samples of Seyfert galaxies with redshifts less than 0.05. The ratio of elliptical host galaxies in the low-redshift Seyfert 1 galaxies is 19\% and that in the low-redshift Seyfert 2 is 14\%. This shows that there are fewer elliptical Seyfert galaxies in low redshifts because there are fewer nearby elliptical galaxies. This explains why people found much fewer elliptical Seyferts in the earlier studies \citep[e.g.,][]{Moles95}.

\begin{figure}
\centering
\includegraphics[width=0.93\textwidth]{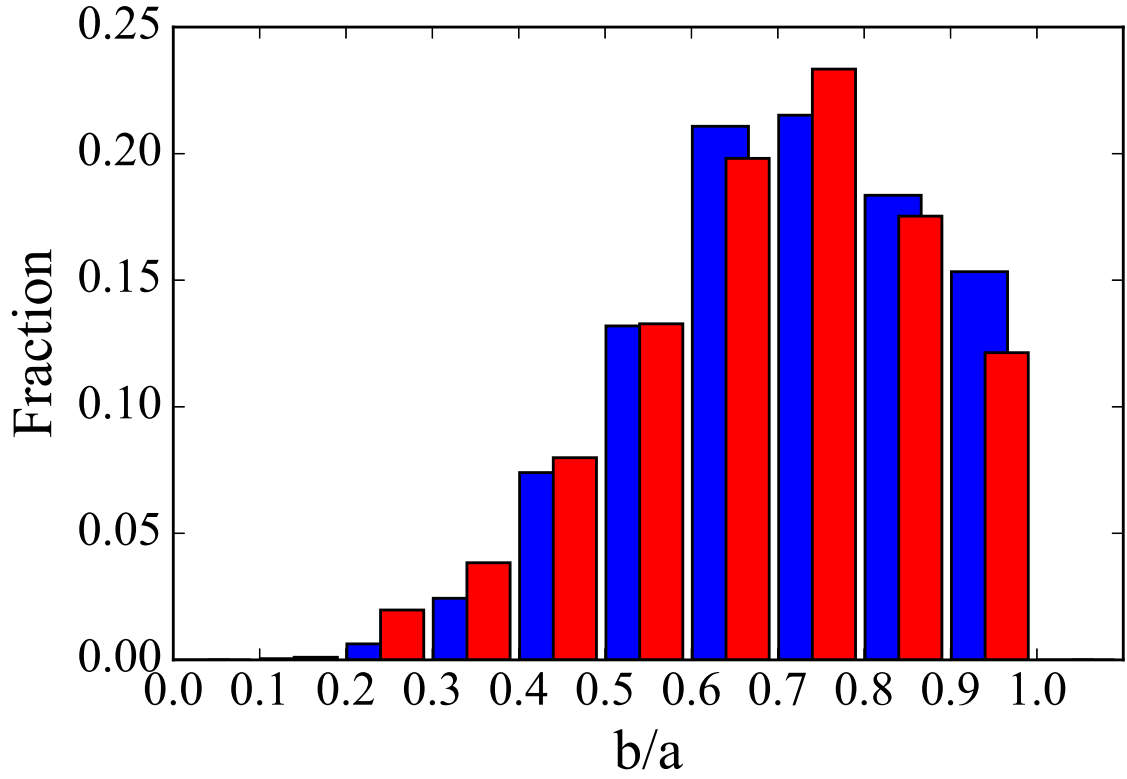} 
\caption{Distributions of axis ratio b/a for Seyfert 1 and Seyfert 2 galaxies with {\it FracDev}$=1$. Blue: Seyfert 1 galaxies. Red: Seyfert 2 galaxies}
\label{s1s2ba}
\end{figure}

The different bulge distributions of Seyfert 1 and Seyfert 2 might be related to galaxy evolution.
\citet{Koulouridis13} indicated that the neighbors of Seyfert 2 galaxies are more ionized than the neighbors of Seyfert 1 galaxies;  they proposed that there is an evolution sequence between Seyfert galaxies. Seyfert 2 galaxies begin from an interaction galaxies and finally transform into Seyfert 1 galaxies. The other factor to cause different bugle distributions of Seyfert 1 and Seyfert 2 galaxies might be related to ISM/orientation of host galaxy. The obscuration of dust in the host galaxy of Seyfert galaxies might affect the identification of the different Seyfert types \citep{Lagos11}. We also test whether our sample is isotropic to the viewing angle of the host galaxy by comparing the axis ratio b/a of the Seyfert galaxies with {\it FracDev} $=1$. The results of axis ratio of Seyfert 1 and Seyfert 2 galaxies are shown in Fig.~\ref{s1s2ba}. We found that both Seyfert 1 and Seyfert 2 galaxies have peak at b/a$=$ 0.6 $-$ 0.7. This result indicates that our Seyfert sample is independent of the viewing angle. It has been arguing that observed type 1 and type 2 AGNs are actually drawn from different distributions of covering factors of the AGNs \citep[e.g.,][]{Elitzur12}.
Therefore, we expect that intrinsic difference should exist in our selected samples of Seyfert 1 and Seyfer 2. However, we note that we are comparing the host of galaxies of the AGNs but not the intrinsic properties of the AGNs; in other words, any isotropic biases in selection should not affect our results about the host galaxies, unless the properties of AGNs were closely related to the host galaxies, which again is not consistent with the unification model. 

\citet{Rutkowski13} found that the host galaxy morphologies of Seyfert 1 and Seyfert 2 are similar. We note that \citet{Rutkowski13} only chose the ``Face-on'' source in order to classify dust features in the cores of their Seyfert galaxies. We found that they tended to select much more disk galaxies than the parent population we have. The similar distribution in the host morphology of Seyfert 1 and Seyfert 2 might be caused by the pre-selection of the ``Face-on'' galaxies, which might exclude most of the elliptical galaxies. Besides, the number of the Seyfert galaxies (30 Seyfert 1 and 53 Seyfert 2) in \citet{Rutkowski13} is too small to distinct the distributions significantly. Although the distributions of host morphology of Seyfert 1 and Seyfert 2 galaxies in the samples of \citet{Rutkowski13} are not distinguishable, they still found that the distributions of the core morphologies of Seyfert 1 and Seyfert 2 galaxies are different.

Besides, \citet{Villarroel14} found the average color of neighbor galaxies around type-I AGN are redder than that of the neighbors around type-II AGN; they also found that the host galaxy morphology of  an AGN depends on the AGN type and the presence of a neighbor. Our results show that the host galaxies of the Seyfert 1 galaxies are dominated by galaxies with large bulge ratios whereas Seyfert 2 galaxies show host galaxies with relatively smaller bulge ratios, suggesting that different Seyferts might appear in the different evolution stages of galaxy evolution.

\begin{figure}
\begin{center}
\includegraphics[width=\columnwidth]{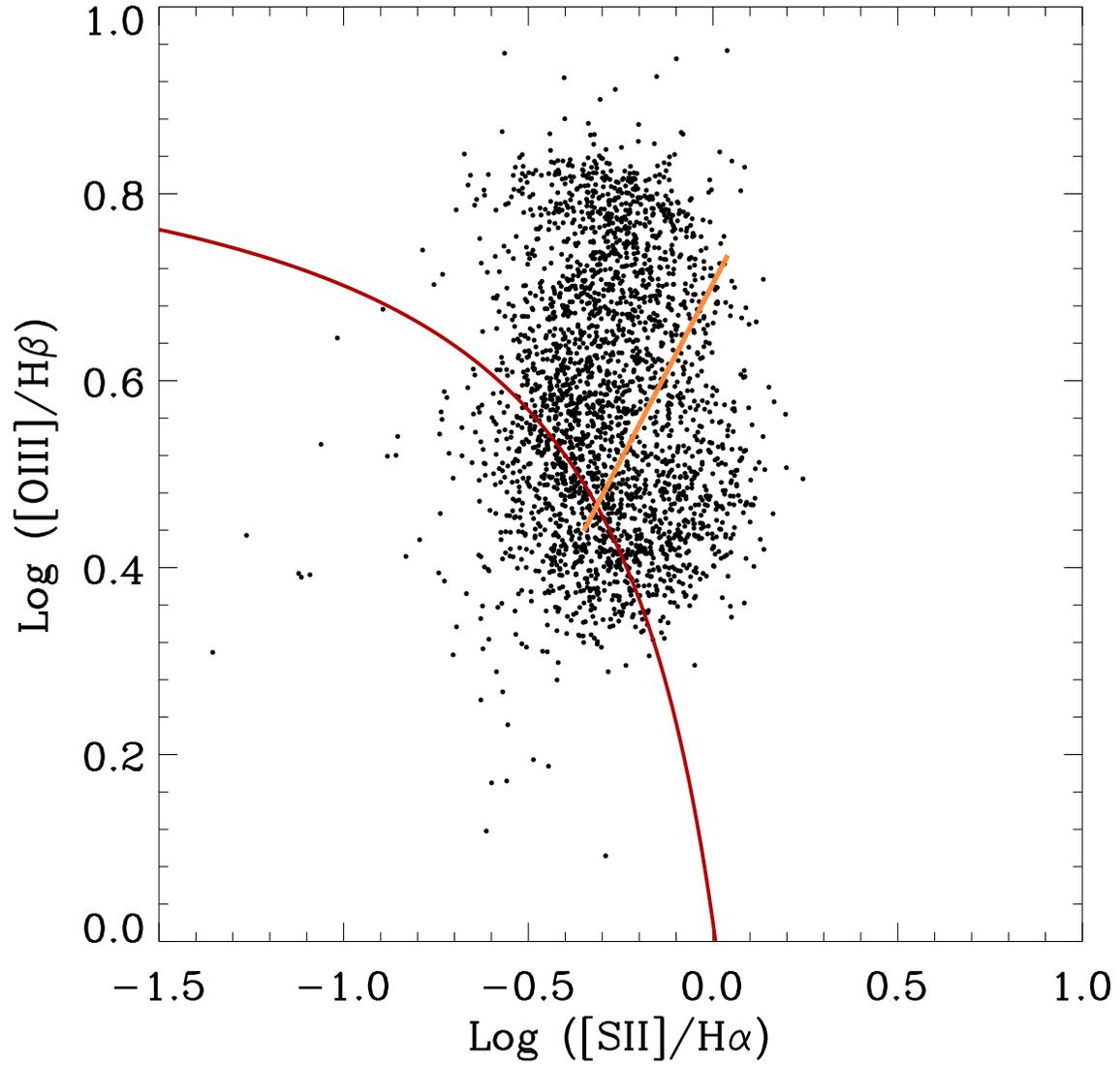}
\caption{Distribution of our Seyfert 2 galaxies on the BPT diagram. The red line represents the definition of the starburst limit of \citet{Kewley01}. The orange line represents the Seyfert-LINER line of \citet{Kewley06}.}
\label{s2bpt}
\end{center}
\end{figure}

\begin{figure}
\begin{center}
\includegraphics[width=\columnwidth]{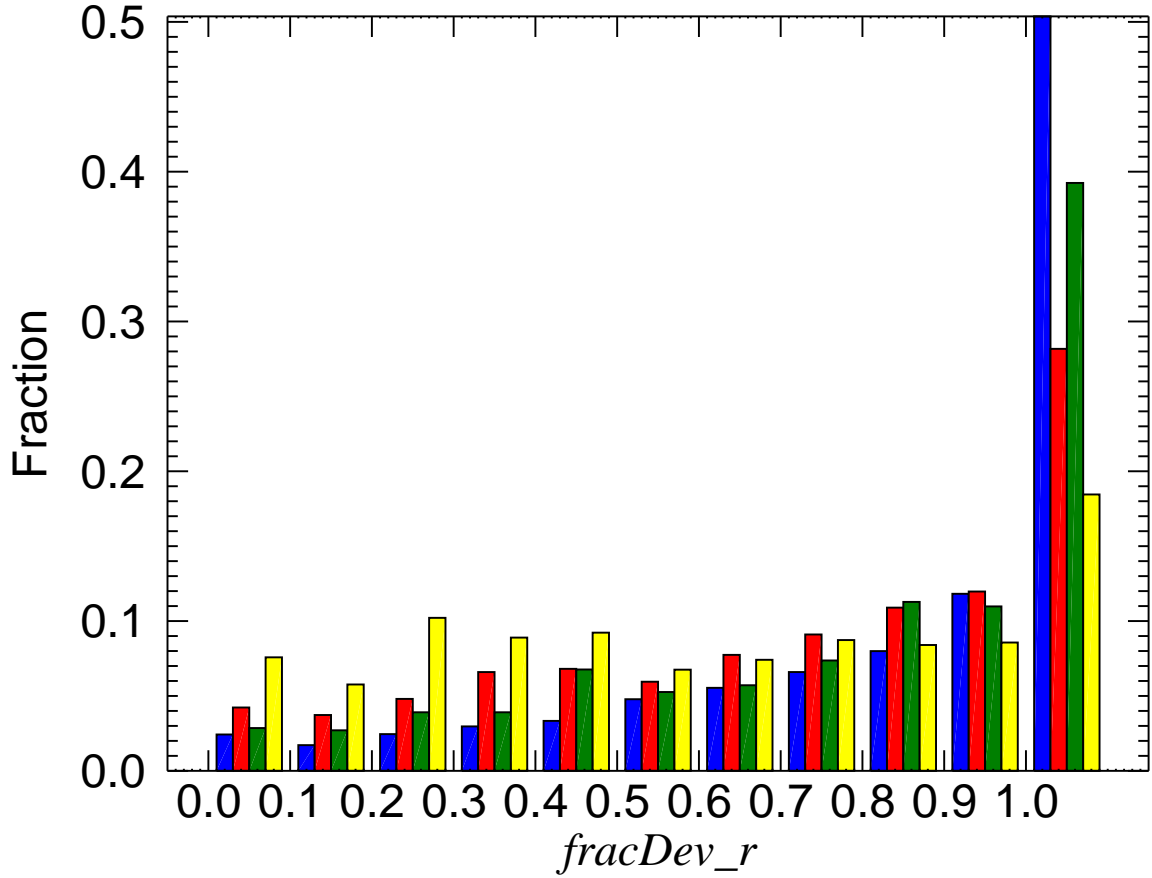}
\caption{{\it{FracDev}} distributions for our Seyfert samples. Blue color represents the Seyfert 1 galaxies. Red color represents the new-defined Seyfert 2 galaxies. Green color represents the LINERs. Yellow color represents the star-forming galaxies.}
\label{s1s2Ls}
\end{center}
\end{figure}

\begin{figure}
\begin{center}
\includegraphics[width=\columnwidth]{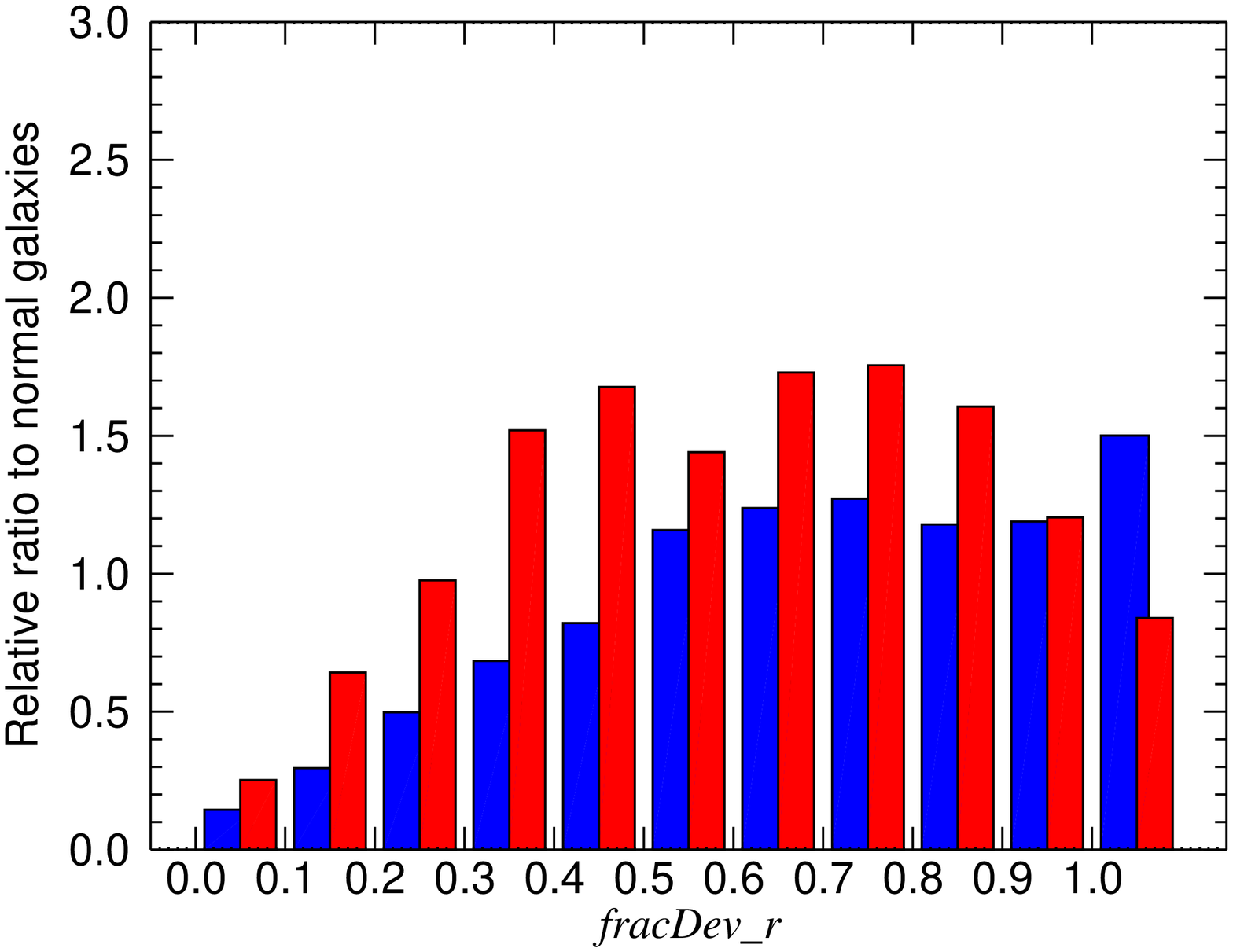}
\caption{Ratios of Seyfert galaxies to all SDSS galaxies with z $<$ 0.2 and $M_{\mathrm{r}} <$ -19. Blue color represents the Seyfert 1 galaxies. Red color represents the new-defined Seyfert 2 galaxies.}
\label{s1ps2}
\end{center}
\end{figure}

We selected our Seyferts samples from the V\'{e}ron catalog, which compiled all AGN sources from observations and literature. However, the definitions of the Seyfert 2 galaxies might be slightly different in different observations. We obtained the photometric and spectral data of these Seyfert galaxies from SDSS \citep{Brinchmann04,Tremonti04} and used the BPT diagram \citep{Baldwin81} to define a more rigorous sample of Seyfert 2 galaxies to check whether our results are affected by the contamination of starburst galaxies and LINERs in the Seyfert 2 sample. We used the definitions of \citet{Kewley01} and \citet{Kewley06} to divide our Seyfert 2 galaxies into three sub-samples: new rigorously defined Seyfert 2, LINERs and starburst galaxies and compare the {\it{FracDev}} distributions of these sub-samples with the Seyfert 1 galaxies. Fig.~\ref{s2bpt} show that some of our original Seyfert 2 galaxies are locating in the LINER and starburst regions.  Fig.~\ref{s1s2Ls} shows the {\it{FracDev}} distributions of the new rigorously defined  Seyfert 2, LINERs, starburst galaxies, and Seyfert 1 galaxies. We found that the {\it{FracDev}} distributions of the new-defined Seyfert 2 and the Seyfert 1 galaxies are still very different and a Kolmogorov-Smirnov test shows that the probability for the two distributions drawn from the same population is $P = 0.00$ 0 with K-S statistic (D) $=0.41$. We replotted the relative ratios of the new-defined Seyfert 2 and Seyfert 1 to all SDSS galaxies with different {\it{FracDev}} values in Fig.~\ref{s1ps2} and we found that the distribution pattens are still similar to Fig.~\ref{s1s2dng}. These results  indicate that the different morphologies of Seyfert 1 and Seyfert 2 galaxies are not caused by contamination of LINERs or starburst galaxies in the Seyfert 2 sample. Besides, we found that the {\it{FracDev}} distributions of the Seyfert 1, the new-defined Seyfert 2 and the LINERs all show distinct peak values at {\it{FracDev}}$=$1, indicating that AGN type galaxies are more bulge-dominant; on the other hand, the {\it{FracDev}} values of the starburst galaxies distribute more evenly.

From our results, we found that the Seyfert 1 and Seyfert 2 galaxies have different host galaxy morphology distributions; Seyfert 2 galaxies have relatively more late-type host galaxies than Seyfert 1 have. Different types of galaxies are expected to have different formation history and physical process in galaxy evolution. The different distributions of the host galaxy morphology suggests that the different types of Seyfert galaxies might be related to the processes of galaxy formation. This indicates that the differences between the Seyfert 1 and Seyfert 2 galaxies are not only affected by the viewing angles but also related to the formation process of their host galaxies.

\subsection{Summary}
{\textmd{We found that Seyfert 1 and Seyfert 2 galaxies have different distributions for their host galaxy morphology. Seyfert 1 galaxies show more bulge-dominated host galaxies and Seyfert 2 galaxies show more disk contributions in their host galaxies. The traditional unification model could not explain these results. This means that the orientation of torus is not the only reason for different types of AGNs. The properties of the host galaxies of AGNs need to be taken into account in an AGN unification model.}

\section*{Acknowledgements}

This work is supported by the Ministry of Science and Technology of Taiwan (grant MOST 106-2119-M-008-016). We thank L. Kewley, P.~C.~Yu, M.~C.~Tsai, Z.~Y.~Chen, J.~C.~Huang, and Y.~L.~Chang for discussion and comments. 

Funding for SDSS-III has been provided by the Alfred P. Sloan Foundation, the Participating Institutions, the National Science Foundation, and the U.S. Department of Energy Office of Science. The SDSS-III web site is http://www.sdss3.org/. SDSS-III is managed by the Astrophysical Research Consortium for the Participating Institutions of the SDSS-III Collaboration including the University of Arizona, the Brazilian Participation Group, Brookhaven National Laboratory, Carnegie Mellon University, University of Florida, the French Participation Group, the German Participation Group, Harvard University, the Instituto de Astrofisica de Canarias, the Michigan State/Notre Dame/JINA Participation Group, Johns Hopkins University, Lawrence Berkeley National Laboratory, Max Planck Institute for Astrophysics, Max Planck Institute for Extraterrestrial Physics, New Mexico State University, New York University, Ohio State University, Pennsylvania State University, University of Portsmouth, Princeton University, the Spanish Participation Group, University of Tokyo, University of Utah, Vanderbilt University, University of Virginia, University of Washington, and Yale University.

This research has made use of the NASA/IPAC Extragalactic Database (NED) which is operated by the Jet Propulsion Laboratory, California Institute of Technology, under contract with the National Aeronautics and Space Administration.






\clearpage



\clearpage

\clearpage




\end{document}